\documentclass[journal]{IEEEtran}

\usepackage{color}
\ifCLASSINFOpdf
  \usepackage[pdftex]{graphicx}
  \DeclareGraphicsExtensions{.pdf,.jpeg,.png}
\else
\fi
\usepackage[cmex10]{amsmath}
\usepackage{amsfonts}
\usepackage{cite}
\usepackage{url}

\newcommand{\J}{\mathrm{j}}                 

\providecommand{\D}{\,\mathrm{d}}           
\providecommand{\M}[1]{\mathbf{#1}}         
\newcommand{\OP}[1]{{\mathcal{#1}}}         

\providecommand{\herm}{\mathrm{H}}

\providecommand{\Prad}{P_\mathrm{r}}
\providecommand{\Wrad}{W_\mathrm{rad}}
\providecommand{\WSTO}{\widetilde{W}_\mathrm{sto}}

\providecommand{\We}{W_\mathrm{e}}
\providecommand{\Wm}{W_\mathrm{m}}

\providecommand{\Ivec}{\M{I}}

\providecommand{\myradius}{\chi}  
\providecommand{\charnum}{\delta}  

\newcommand\figwidth{9} 

\newcommand{\degree}{\ensuremath{^\circ}}

\begin{document}
\title{Analytical Representation of Characteristic Modes Decomposition}
\author{Miloslav~Capek,~\IEEEmembership{Member,~IEEE,}
        Pavel~Hazdra,~\IEEEmembership{Member,~IEEE,}
        Michal~Masek,
        Vit~Losenicky
\thanks{Manuscript received July XX, XXXX; revised January XX, XXXX.
This work was supported by the project of the Technology Agency of the Czech Republic, No.~TA04010457 and by the Grant Agency of the Czech Technical University in Prague SGS16/226/OHK3/3T/13.}
\thanks{The authors are with the Department of Electromagnetic Field, Faculty of Electrical Engineering, Czech Technical University in Prague, Technicka 2, 16627, Prague, Czech Republic
(e-mail: miloslav.capek@fel.cvut.cz).}
}

%

\markboth{Journal of \LaTeX\ Class Files,~Vol.~PP, No.~X, January~2017}%
{Capek \MakeLowercase{\textit{et al.}}: Characteristic Modes and Their Electromagnetic Properties}
%


\maketitle

\begin{abstract}
Aspects of the theory of characteristic modes, based on their variational formulation, are presented and an explicit form of a related functional, involving only currents in a spatial domain, is derived. The new formulation leads to deeper insight into the modal behavior of radiating structures as demonstrated by a detailed analysis of three canonical structures: a dipole, an array of two dipoles and a loop, cylinder and a sphere. It is demonstrated that knowledge of the analytical functional can be utilized to solve important problems related to the theory of characteristic modes decomposition such as the resonance of inductive modes or the benchmarking of method of moments code.
\end{abstract}


%
\IEEEpeerreviewmaketitle

\section{Introduction}
\label{Intro}
\IEEEPARstart{T}{he} theory of characteristic modes (CMs), formally developed by Garbacz \cite{Garbacz_TCMdissertation} and Harrington and Mautz \cite{HarringtonMautz_TheoryOfCharacteristicModesForConductingBodies}, has become very popular in recent years as this theory constitutes a general approach to characterizing the modal resonant behavior of arbitrarily shaped antennas and scatterers \cite{MartaEva_TheTCMRevisited}. In its original form, which is considered here, the CM assumes perfect electric conductors (PEC) in a vacuum. Academic interest and a number of publications dealing with CMs continue to grow. However, most papers focus only on the application character, such as \cite{DaviuFabresGalloBataller_DesignOfAMultimodeMIMOantennaUsingCM,Obeidat_TCMdissertation,Adams_CharacteristicModesForImpedanceMatching}. Excluding the first attempt to summarize CMs in a book \cite{ChenWang_CharacteristicModesWiley}, there are also related chapters to be found in older books \cite{Bladel_ElmagField} and \cite{Mittra_TopicsInAppliedPhysics}.

This paper briefly reviews characteristic mode decomposition and what constitutes the necessary theoretical background. An analytical form of the functional, composed of reactive and radiated power, is derived, based on previous research \cite{Vandenbosch_ReactiveEnergiesImpedanceAndQFactorOfRadiatingStructures}, \cite{CapekJelinekHazdraEichler_MeasurableQ}. This relation has to be satisfied for each mode but is not restricted to the characteristic basis. Hence, it is possible to specify arbitrary current distribution (the CM can be predicted, see \cite{WuSu_AbroadbandModelOfTheCMForRectangularPlates}) and compare it with real CMs. Based on this result, properties of canonical shapes are investigated, including inductive modes. Analogically, if the modes are analytically known, they can be substituted into a derived functional instead of using an approximative solution given by the numerical spectral decomposition of an underlying operator.

\section{Derivation of the functional}
\label{TCM}

Based on previous work by Garbacz \cite{Garbacz_TCMdissertation}, Harrington \cite{HarringtonMautz_TheoryOfCharacteristicModesForConductingBodies} reduced the CMs into the following generalized eigenvalue problem (GEP, \cite{Sagan_BoundaryEigenvalueProblemsInMathematicalPhysics})
\begin{equation}
\OP{X} \left(\boldsymbol{J}_{n}\right) = \lambda_{n} \OP{R} \left(\boldsymbol{J}_{n} \right),
\label{TCM_eq4}
\end{equation}
where $\OP{R}$ and $\OP{X}$ are real and symmetric operators forming the impedance operator
\begin{equation}
\OP{Z} \left(\boldsymbol{J}_{n}\right)=\OP{R} \left(\boldsymbol{J}_{n}\right)+\J\OP{X} \left(\boldsymbol{J}_{n}\right) = \boldsymbol{n}_0 \times \left(\J \omega \boldsymbol{A} + \nabla\phi \right),
\label{TCM_eq4A}
\end{equation}
$\boldsymbol{A}$ and $\phi$ are corresponding magnetic and electric time-harmonic potentials in Lorenz gauge \cite{Jackson_ClassicalElectrodynamics}, $\boldsymbol{J}_{n}$ is the modal current density, and $\boldsymbol{n}_0$ is the unit vector tangential to the PEC boundary of a radiator. The continuous operator $\OP{Z}$ is usually discretised by the method of moments (MoM, \cite{Harrington_FieldComputationByMoM}), utilizing a proper set of basis functions 
\begin{equation}
\label{TCM_storedEnergy10}
\boldsymbol{J}_n \left(\boldsymbol{r}\right) \approx \sum\limits_{m=1}^N I_{mn} \boldsymbol{f}_m \left(\boldsymbol{r}\right),
\end{equation}
where $I_{mn}$ are (modal) expansion coefficients and $\boldsymbol{f}_m \left(\boldsymbol{r}\right)$ are frequency-independent basis functions, e.g., RWG basis functions \cite{RaoWiltonGlisson_ElectromagneticScatteringBySurfacesOfArbitraryShape}. Consequently, the MoM procedure leads to an impedance matrix \mbox{$\mathbf{Z} = \mathbf{R} + \J \mathbf{X}$}, which is the discrete representation of the analytical operator $\OP{Z}$. Finally, the CMs can be defined in (common) algebraic form \cite{HarringtonMautz_TheoryOfCharacteristicModesForConductingBodies}
\begin{equation}
\mathbf{X} \Ivec_n = \lambda_{n} \mathbf{R} \Ivec_n,
\label{TCM_eq4B}
\end{equation}
which is, in comparison to (\ref{TCM_eq4}), numerically solvable for an arbitrary radiator since it is based on real and symmetric matrices of size $N\times N$, where $N$ is the number of basis functions.

The solution of the GEP produces the characteristic basis $\{ \boldsymbol{J}_n, \lambda_n \}$ of eigencurrents $\boldsymbol{J}_n$ and associated eigenvalues $\lambda_n$ and, due to the properties of the impedance matrix, all eigenvalues are real with all eigencurrents equiphasal (they can also be selected as real, \cite{HarringtonMautz_ComputationOfCharacteristicModesForConductingBodies}). Furthermore, the CMs minimize the ratio of the net reactive power $\omega \left(\Wm - \We\right)$ to radiated power $\Prad$. Note that the extremal value of radiated to stored power is considered for the basis as a whole.

It is known \cite{ChenWang_CharacteristicModesWiley} that the GEP (\ref{TCM_eq4}) minimizes a power functional\footnote{Through this paper, the following notation is used $\langle \boldsymbol{f}, \boldsymbol{g} \rangle = \int_{\Omega} \boldsymbol{f}^*\cdot\boldsymbol{g}~\mathrm{d}\Omega$ and $\langle \boldsymbol{f}, \boldsymbol{g} \rangle_\mathrm{r} = \int_{\Omega} \boldsymbol{f}\cdot\boldsymbol{g}~\mathrm{d}\Omega$.}
\begin{equation}
\mathcal{F}\left(\boldsymbol{J}_n\right) = \frac{\langle \boldsymbol{J}_{n}, \OP{X} \boldsymbol{J}_{n} \rangle}{\langle \boldsymbol{J}_{n}, \OP{R} \boldsymbol{J}_{n} \rangle} = \frac{2\omega(W_\mathrm{m}^{n}-W_\mathrm{e}^{n})}{P_\mathrm{r}^{n}}=\lambda_{n},
\label{TCM_eq5}
\end{equation}
where $W_\mathrm{m}^{n}$ and $W_\mathrm{e}^{n}$ are modal magnetic and electric potentials-based energies, defined here as
\begin{equation}
W_\mathrm{m}^{n} = \frac{1}{2} \Re \int\limits_V \boldsymbol{A}\cdot\boldsymbol{J}_n^\ast \D{V},
\label{TCM_eq6A}
\end{equation}
\begin{equation}
W_\mathrm{e}^{n} = \frac{1}{2} \Re \int\limits_V \varphi \rho_n^\ast \D{V},
\label{TCM_eq6B}
\end{equation}
with $P_\mathrm{r}^{n}$ as modal radiated power which is commonly normalized as $P_\mathrm{r}^{n}=1\,$W. It should be noted that energies (\ref{TCM_eq6A}) and (\ref{TCM_eq6B}) are not equal to true electric (\mbox{$\int_V \epsilon \|\boldsymbol{E}\|^2\D{V}/2$}) and magnetic (\mbox{$\int_V \mu \|\boldsymbol{H}\|^2\D{V}/2$}) energy \cite{Carpenter_ElectromagneticEnergyAndPowerInTermsOfChargesAndPotentialsInsteadOfFields}. However, a clear advantage of (\ref{TCM_eq6A}) and (\ref{TCM_eq6B}) is that they can be calculated easily and directly from the (characteristic) currents if they are prescribed analytically or calculated numerically. The paradigm used, and its further extension towards the stored energy, is briefly discussed in Section~\ref{StoredEnergy}.

A particular form of the above mentioned functional (\ref{TCM_eq5}), established directly for the sources (currents/charges) on the antenna, is derived using (\ref{TCM_eq4A}) and it reads
\begin{equation}
\mathcal{F} \big( \boldsymbol{J}_{n} \big) = \frac{\langle \boldsymbol{J}_{n}, \OP{X} \boldsymbol{J}_{n} \rangle}{\langle \boldsymbol{J}_{n}, \OP{R} \boldsymbol{J}_{n} \rangle} = -
\frac{\displaystyle\Re\int\limits_{V}\left(\boldsymbol{A}\cdot\boldsymbol{J}_{n}^*-\phi\rho_{n}^*\right)\mathrm{d}V}{\displaystyle\Im\int\limits_{V}\left(\boldsymbol{A}\cdot\boldsymbol{J}_{n}^*-\phi\rho_{n}^*\right)\mathrm{d}V},
\label{TCMfcnAJ}
\end{equation}
where $V$ is the volume of an antenna and $\rho_{n}$ is the charge density. Inserting the continuity equation \cite{Jackson_ClassicalElectrodynamics}, \mbox{$\rho=-\nabla\cdot\boldsymbol{J}/\mathrm{j}\omega$}, the functional involves only currents and reads
\begin{equation}
\label{TCM2_eq8}
\mathcal{F} \big( \boldsymbol{J}_{n} \big) =
\frac{\displaystyle\int\limits_{V} \!\! \int\limits_{V '} \mathcal{J}\big(\boldsymbol{J}_{n}\big) \frac{\cos (k R)}{R} \, \mathrm{d} V' \, \mathrm{d} V}{ \displaystyle\int\limits_{V} \!\! \int\limits_{V '} \mathcal{J}\big(\boldsymbol{J}_{n}\big) \frac{\sin (k R)}{R} \, \mathrm{d} V' \, \mathrm{d} V} = \kappa_n,
\end{equation}
where $\mathcal{J}\big(\boldsymbol{J}_{n}\big)=\big( k^2 \boldsymbol{J}_{n} (\boldsymbol{r}) \cdot \boldsymbol{J}_{n}^{*} (\boldsymbol{r} ') - \nabla \cdot \boldsymbol{J}_{n} (\boldsymbol{r}) \nabla ' \cdot \boldsymbol{J}_{n}^{*} (\boldsymbol{r} ') \big)$,  $R=|\boldsymbol{r}-\boldsymbol{r}'|$ is Euclidean distance, $k$ is the wavenumber and $\kappa_{n}$ is the Rayleigh quotient \cite{Stewart_Sun_MatrixPerturbationTheory}, which is equal to characteristic number $\lambda_{n}$ when the true characteristic current $\boldsymbol{J}_{n}$ enters into (\ref{TCM2_eq8}).

Thanks to the ``source'' formulation (\ref{TCM2_eq8}), arbitrary current distribution can be studied and its properties with true CMs can be compared. This formulation extends the understanding of the original definition in \cite{HarringtonMautz_TheoryOfCharacteristicModesForConductingBodies}, since, as will be shown later, we can study the separated components\footnote{In the numerator, the net reactive power may be further split into its ``current'' and ``charge'' parts to express the modified magnetic and electric energies separately. For more details see \cite{Vandenbosch_ReactiveEnergiesImpedanceAndQFactorOfRadiatingStructures, HazdraCapekEichler_CommentsToGuy1, Vandenbosch_Reply2Comments, Gustaffson_StoredElectromagneticEnergy_PIER}.} of (\ref{TCM2_eq8}).

It is important to stress that the functional is minimized by characteristic currents, i.e. solutions of (\ref{TCM_eq4}). Such a (eigen) basis maximizes the radiated power and minimizes the net reactive power, indicating external resonances of the radiator. Hence, the extremum of (\ref{TCM2_eq8}) is given by characteristic basis $\{\boldsymbol{J}_n\}$ with associated eigenvalues $\lambda_n$.

An exact analytical solution for characteristic currents is exceedingly complicated with only two bodies of finite extent already known, one of them being a spherical shell \cite{EvaDaviu_TCMdissertation}. However, the expression (\ref{TCM2_eq8}) permits the definition of an arbitrary current distribution $\widetilde{\boldsymbol{J}}$ without the necessity of numerically computing the impedance matrix $\mathbf{Z}$ and its decomposition in (\ref{TCM_eq4B}). In addition, if we analytically try to test a basis $\widetilde{\boldsymbol{J}}$ that is similar to the true CM basis, we can precisely analyze its behaviour and estimate how close the selected current distribution is to the optimal solution \cite{WuSu_AbroadbandModelOfTheCMForRectangularPlates}.

\subsection{Relation Between Characteristic Modes and Stored Enegy}
\label{StoredEnergy}

There is an interesting relationship between decomposition into CMs and the evaluation of the modified stored electromagnetic energy, proposed by Vandenbosch in \cite{Vandenbosch_ReactiveEnergiesImpedanceAndQFactorOfRadiatingStructures} as
\begin{equation}
\label{TCM_storedEnergyN1}
\WSTO = \frac{1}{4}\left\langle \boldsymbol{J}, \frac{\partial \OP{X}}{\partial\omega} \boldsymbol{J} \right\rangle,
\end{equation}
in which the structure of $\OP{X}$ is obvious from (\ref{TCM_eq4A}) and $\boldsymbol{J}$ is the current density which, in the context of this paper, can be composed as
\begin{equation}
\label{TCM_storedEnergyN3}
\boldsymbol{J} \approx \sum\limits_n \alpha_n \boldsymbol{J}_n,
\end{equation}
where $\alpha_n$ is given in \cite{HarringtonMautz_TheoryOfCharacteristicModesForConductingBodies}. If (\ref{TCM_storedEnergy10}) is substituted, (\ref{TCM_storedEnergyN1}) can be represented in a useful matrix form as proposed by Gustafsson et al. \cite{CismasuGustafsson_FBWbySimpleFreuqSimulation}
\begin{equation}
\label{TCM_storedEnergyN2}
\WSTO \approx \frac{1}{4} \Ivec^\herm \frac{\partial \mathbf{X}}{\partial\omega} \Ivec
\end{equation}
and anticipated much earlier by Harrington and Mautz \cite{HarringtonMautz_ControlOfRadarScatteringByReactiveLoading}.

It is argued in \cite{HarringtonMautz_ControlOfRadarScatteringByReactiveLoading} that in the vicinity of $n$th modal resonances the quality factor~$Q_n$, defined as
\begin{equation}
\label{TCM_storedEnergyN10}
Q_n = \frac{\omega}{2} \frac{\partial\lambda_n}{\partial\omega},
\end{equation}
is approximately equal to the quality factor rigorously derived by Vandenbosch and later reformulated by Gustafsson, i.e.,
\begin{equation}
\label{TCM_storedEnergyN11}
Q_{\mathbf{X},n} = \frac{\omega \WSTO^n}{P_\mathrm{r}^n} \approx Q_n.
\end{equation}
The argumentation is based on the assumption that the dominant frequency variation is due to the imaginary part of the impedance matrix \cite{HarringtonMautz_ControlOfRadarScatteringByReactiveLoading}. Interestingly, the relationship between these two quality factors\footnote{Please, keep in mind that there are number of quality factor~$Q$ definitions through the literature with possible different meaning \cite{VolakisChenFujimoto_SmallAntennas}.} can be expressed rigorously as
\begin{equation}
\label{TCM_storedEnergyN12}
Q_n = Q_{\mathbf{X},n} - \lambda_n Q_{\mathbf{R},n},
\end{equation}
in which $Q_{\mathbf{R},n}$ is defined in the same way as $Q_{\mathbf{X},n}$, although $\mathcal{R}$ or $\mathbf{R}$ is used instead of $\mathcal{X}$ or $\mathbf{X}$. For the exact derivation of (\ref{TCM_storedEnergyN12}), see Appendix~\ref{app1}. Moreover, the above-mentioned assumption is not needed since the equality \mbox{$Q_n = Q_{\mathbf{X},n}$} is based on definition (\ref{TCM_eq5}) where the eigenvalues are zero at the modal resonances.

Equality between (\ref{TCM_storedEnergyN10}), (\ref{TCM_storedEnergyN12}) and (\ref{TCM_storedEnergyN1}), (\ref{TCM_storedEnergyN2}) establishes explicit link between frequency behavior of eigenvalues $\lambda_n$ and modified modal stored energies \cite{CapekHazdraEichler_AMethodForTheEvaluationOfRadiationQBasedOnModalApproach}. This connection is possible thanks to the modal potential-based energies (\ref{TCM_eq6A}), (\ref{TCM_eq6B}), which occur both in definition of eigenvalues (\ref{TCM_eq5}) and in (\ref{TCM_storedEnergyN1}) through (\ref{TCM_cylinder7}).

\section{Elementary Radiators -- Case Studies}
\label{CaseStudies}

In certain (simple) cases the CM basis can be sufficiently approximated by analytical currents. We inspect three canonical examples:
\begin{itemize}
    \item a thin-strip dipole (Section~\ref{CS_dipole}),
    \item two parallel coupled dipoles, separated by distance $h$ with in-phase and out-of-phase modes (Section~\ref{CS_dipoles}),
    \item a loop with uniform mode (Section~\ref{CS_loop}).
\end{itemize}
These examples establish a direct way to understand stationary inductive modes. It will be seen that these fulfil $\nabla \cdot \boldsymbol{J} (\boldsymbol{r})=0$, i.e., they have no charge. Observations denoted in this section introduce material which is to be developed in Section~\ref{Fun_examples}.

\subsection{Thin-strip dipole}
\label{CS_dipole}

Let us consider a thin-strip dipole of length $L$ and width \mbox{$w = L / 100$}. Since the dipole is thin the inductive modes are not considered and the current has to fulfill the Dirichlet boundary condition at its ends. It is significant that the choice of any mode from the basis predestinates the basis, as a whole, as the modes are orthogonal. We consider the natural first-order current basis\footnote{The tilde in $\widetilde{\boldsymbol{J}}_n (z)$ expresses that we insert artificial current, since an exact form of the mode is not known.}

\begin{equation}
\label{TCM_dip1_Ex1}
\mathbf{\widetilde{\boldsymbol{J}}}_n (z) = \mathbf{z}_0 K_0 \delta(y) \sin \Big(\frac{\pi n z}{L}\Big), x \in \left(-\frac{w}{2},\frac{w}{2}\right), z \in (0,L),
\end{equation}
where the surface current density 
\begin{equation}
\label{TCM_loop1_Ex1A}
K_0 = \frac{1}{h}
\end{equation}
is assumed. The divergence of (\ref{TCM_dip1_Ex1}) is
\begin{equation}
\label{TCM_dip1_Ex1B}
\mathbf{z}_0\cdot\frac{\partial \widetilde{\boldsymbol{J}}_n (z)}{\partial z} = K_0 \delta(y) \frac{\pi n}{L} \cos \Big(\frac{\pi n z}{L}\Big).
\end{equation}

\begin{figure}[]
\centering
\includegraphics[width=\figwidth cm]{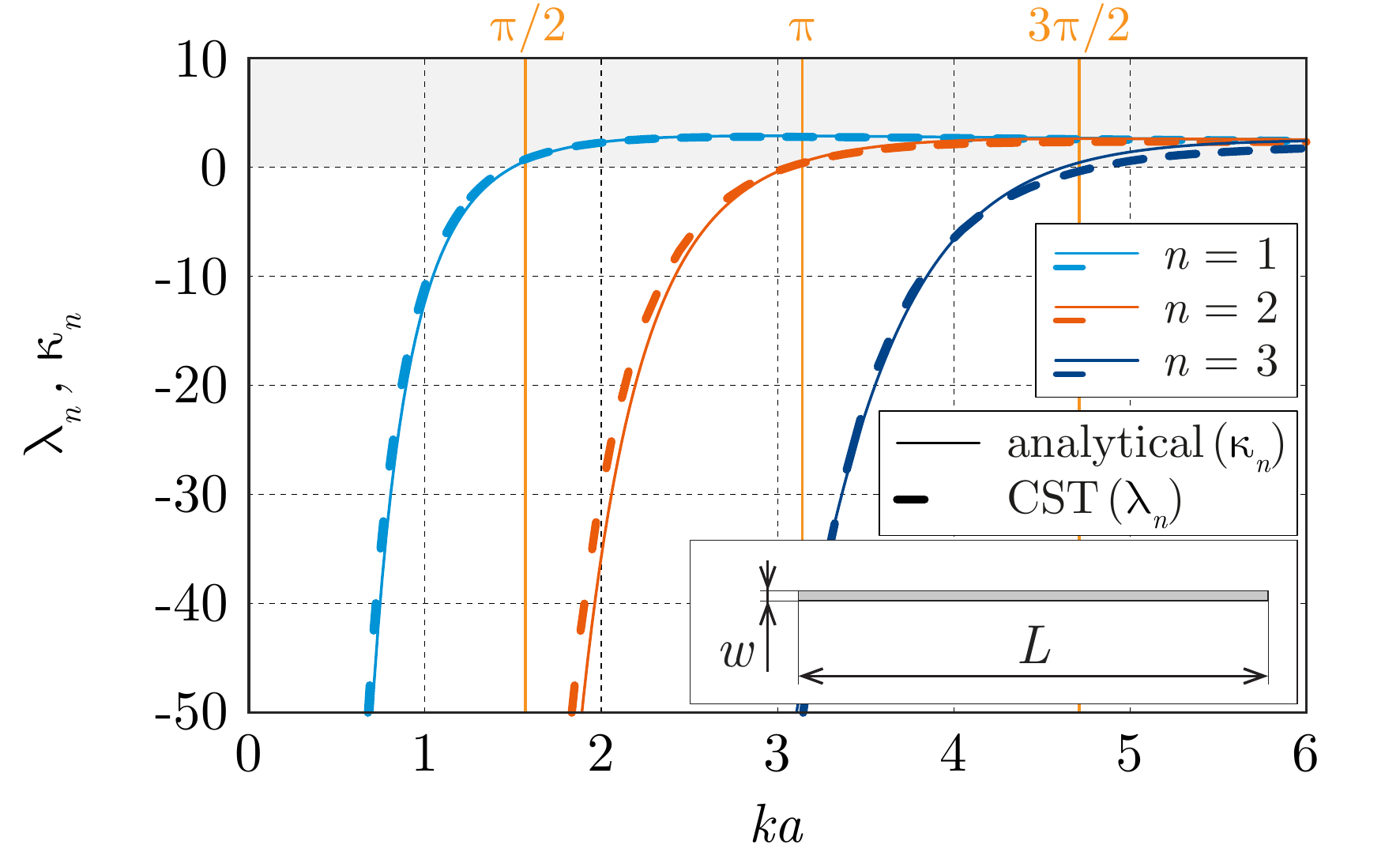}
\caption{The radiation quotients $\kappa_n$ for the first three natural modes of a thin-strip dipole (ratio $L/w=100$ and 962 triangular segments used for numerical calculation) compared to CM eigenvalues $\lambda_n$ from CST-MWS. Resonance of modes occur for \mbox{$ka\cong n\pi/2$} which agree well with theoretical predictions.}
\label{fig_dipole_fig1}
\end{figure}

Due to the complexity\footnote{Following the tedious induced-EMF procedure \cite{JordanBalmain_EMwavesAndRadiatingSystems} for basis (\ref{TCM_dip1_Ex1}), closed form solution to (\ref{TCM2_eq8}) can be found. It is expected (see results for slightly different basis treated in \cite[chapter~14]{JordanBalmain_EMwavesAndRadiatingSystems}) that the results would present similar complexity, not giving additional physical insight.} of (\ref{TCM_dip1_Ex1}), (\ref{TCM_dip1_Ex1}) and (\ref{TCM_dip1_Ex1B}) were inserted in (\ref{TCM2_eq8}) and solved numerically in MATLAB \cite{matlab}. First, three modes, \mbox{$n = \left\{1,2,3\right\}$}, are considered. Figure~\ref{fig_dipole_fig1} shows the $\kappa_{n}$ quotients, together with exact eigenvalues $\lambda_{n}$, obtained by solving (\ref{TCM_eq4}) in CST-MWS software \cite{CST}. A good match is attained, even for such a simple basis (\ref{TCM_dip1_Ex1}).

\begin{figure}[]
\centering
\includegraphics[width=\figwidth cm]{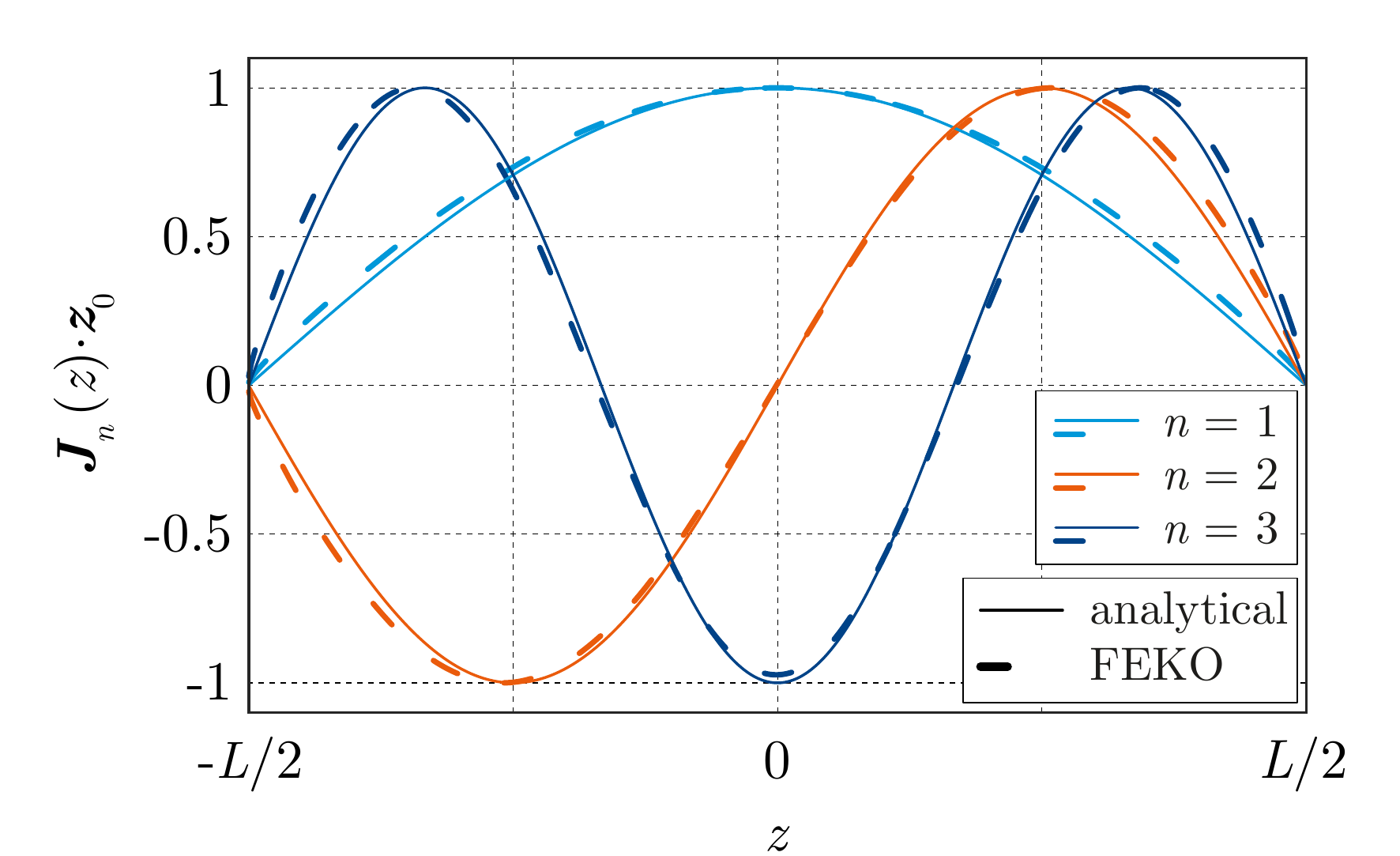}
\caption{Comparison of characteristic modes calculated in FEKO (discretization into 407 linear segments) and analytical current distribution (\ref{TCM_dip1_Ex1}) for the first three modes at resonance on a thin-wire dipole. The amplitude of all three modes is normalized to unity.}
\label{fig_dipole_fig2}
\end{figure}

It can be seen from Fig.~\ref{fig_dipole_fig2} that the agreement between the CM current and its approximation is good, especially for the dominant mode. The analytical current in (\ref{TCM_dip1_Ex1}) is, in fact, exact for a non-radiating 1D resonator, while, in turn, the real CMs maximize radiation and, thus, the shape slightly deviates from the sine basis (\ref{TCM_dip1_Ex1}), \cite{Papas_TheoryOfElectromagneticWavePropagation}.

\subsection{Two thin-strip dipoles}
\label{CS_dipoles}

The next scenario involves two closely spaced collinear thin-strip dipoles with length $L$, separation \mbox{$h = L / 50$} and strip width of \mbox{$w = L/100$}. There are, depending on the actual orientation of currents, two possible basic modes: in-phase and out-of phase. Currents are considered in the form of fundamental distribution $\widetilde{\boldsymbol{J}}_1$ from (\ref{TCM_dip1_Ex1}).

\begin{figure}[]
\centering
\includegraphics[width=\figwidth cm]{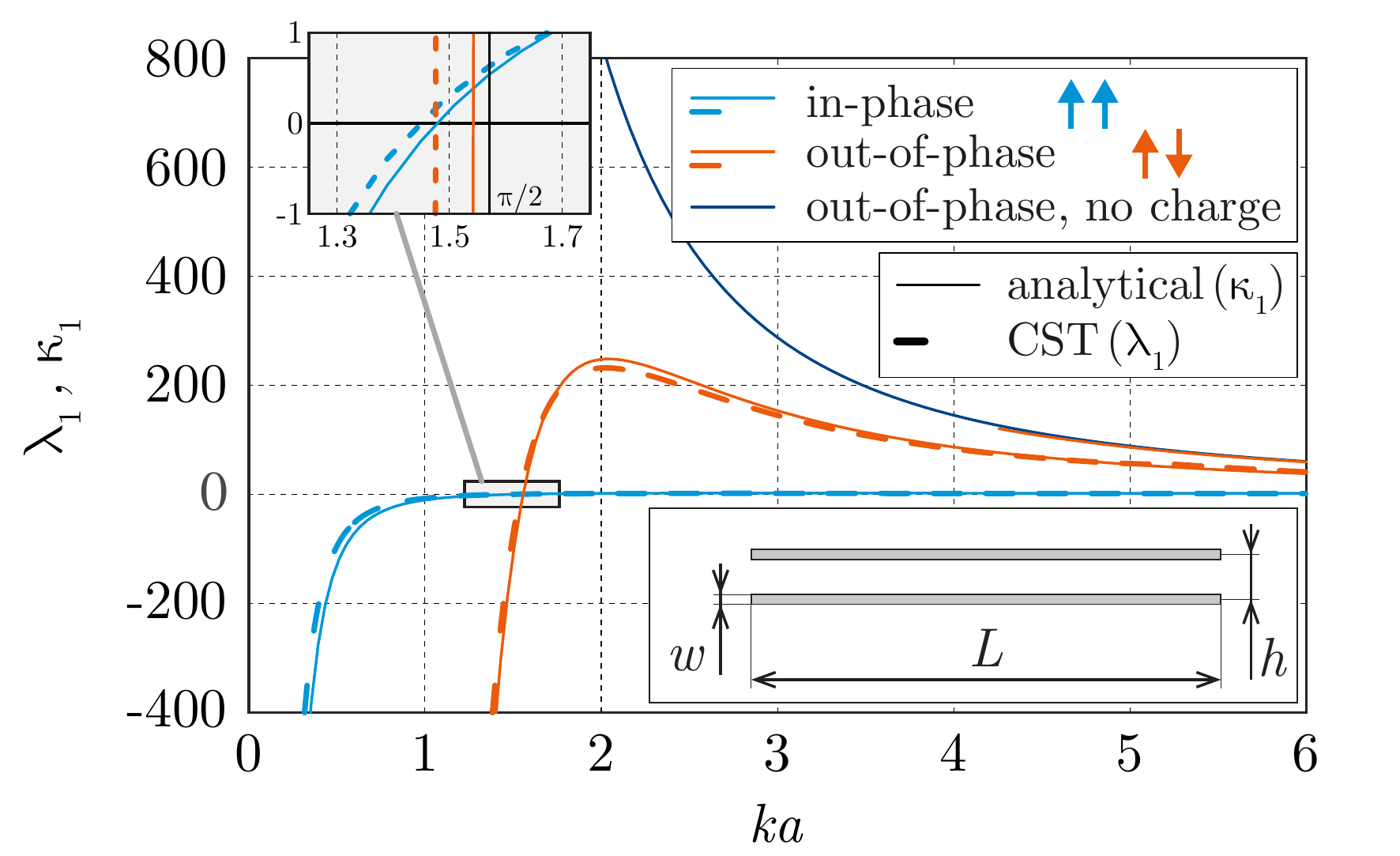}
\caption{The radiation quotients $\kappa_1$ for in-phase, out-of-phase, and testing current with no charge ($\nabla \cdot \widetilde{\boldsymbol{J}}_1 \equiv 0$) of two closely spaced thin wire (\mbox{$L/h = 100$}) dipoles. The detail of the position at which resonance occurs is depicted in the inset. The results are compared with the eigenvalues $\lambda_1$ of the same (thin-strip) structure calculated in CST (except for the testing current with no charge which is artificial), where 1924 triangle elements have been used.}
\label{fig_dipoles_fig1}
\end{figure}

For the in-phase mode \cite{HazdraCapekEichler_RadiationQFactorsOfThinWireDipoleArrangemens}, the course of the $\kappa_1$ quotient (light-blue line at Fig.~\ref{fig_dipoles_fig1}) is similar to that of the dominant mode on a single dipole. It radiates well and the two in-phase currents may be interpreted as one, flowing along a thicker dipole in a manner similar to a folded dipole. This is not the case for the out-of-phase mode, where the radiated power is much lower. Consequently, the orange line in Fig.~\ref{fig_dipoles_fig1} shows extremely steep resonance for this mode. Other properties, especially those regarding radiated Q factors, have been discussed in \cite{CapekHazdraEichler_AMethodForTheEvaluationOfRadiationQBasedOnModalApproach} and analytically treated in \cite{HazdraCapekEichlerMazanek_DipoleRadiationAboveGroundPlane}.

Using (\ref{TCM2_eq8}), it is possible to investigate the hypothetical situation where the currents on the dipoles are out-of-phase but with the charge density eliminated (\mbox{$\nabla \cdot \widetilde{\boldsymbol{J}}_1 \equiv 0$}). It strongly resembles the situation where the ends of the dipoles are connected to form a loop. The dark-blue line in Fig.~\ref{fig_dipoles_fig1} reveals that this mode does not resonate because the ``charge'' part in (\ref{TCM2_eq8}) is missing and the mode, thus, exhibits pure inductive character. In the next section we show that this behaviour is similar to the uniform zero-order mode on a loop.

\subsection{A loop}
\label{CS_loop}

A loop is an elementary radiator on which the uniform (also termed static or inductive) mode with $\nabla\cdot \widetilde{\boldsymbol{J}}_0=0$ exists and its behaviour is similar to the modified out-of-phase mode previously analysed. Current distribution on a thin-wire loop of radius $\myradius$ and height $h = \myradius/100$ is expressed in cylindrical coordinates ($r$,~$\varphi$,~$z$) as
\begin{equation}
\label{TCM_loop1_Ex1}
\widetilde{\boldsymbol{J}}_0 \left(\varphi,r,z\right) = \boldsymbol{\varphi}_0 K_0 \left(z\right) \delta \left(r - \myradius\right), \quad z \in \left(-\frac{h}{2}, \frac{h}{2}\right)
\end{equation}
with surface density (\ref{TCM_loop1_Ex1A}), which simplifies (\ref{TCM2_eq8}) to
\begin{equation}
\label{TCM_loop1_Ex2}
\kappa_0 = \frac{\displaystyle \int\limits_0^{2 \pi} \cos\left( \varphi \right) \frac{\cos \left(k \myradius \varphi\right)}{\myradius \varphi} \, \mathrm{d} \varphi}{\displaystyle \int\limits_0^{2 \pi} \cos\left( \varphi\right) \frac{\sin \left(k \myradius \varphi \right)}{\myradius \varphi}
\, \mathrm{d} \varphi}.
\end{equation}

\begin{figure}[]
\centering
\includegraphics[width=\figwidth cm]{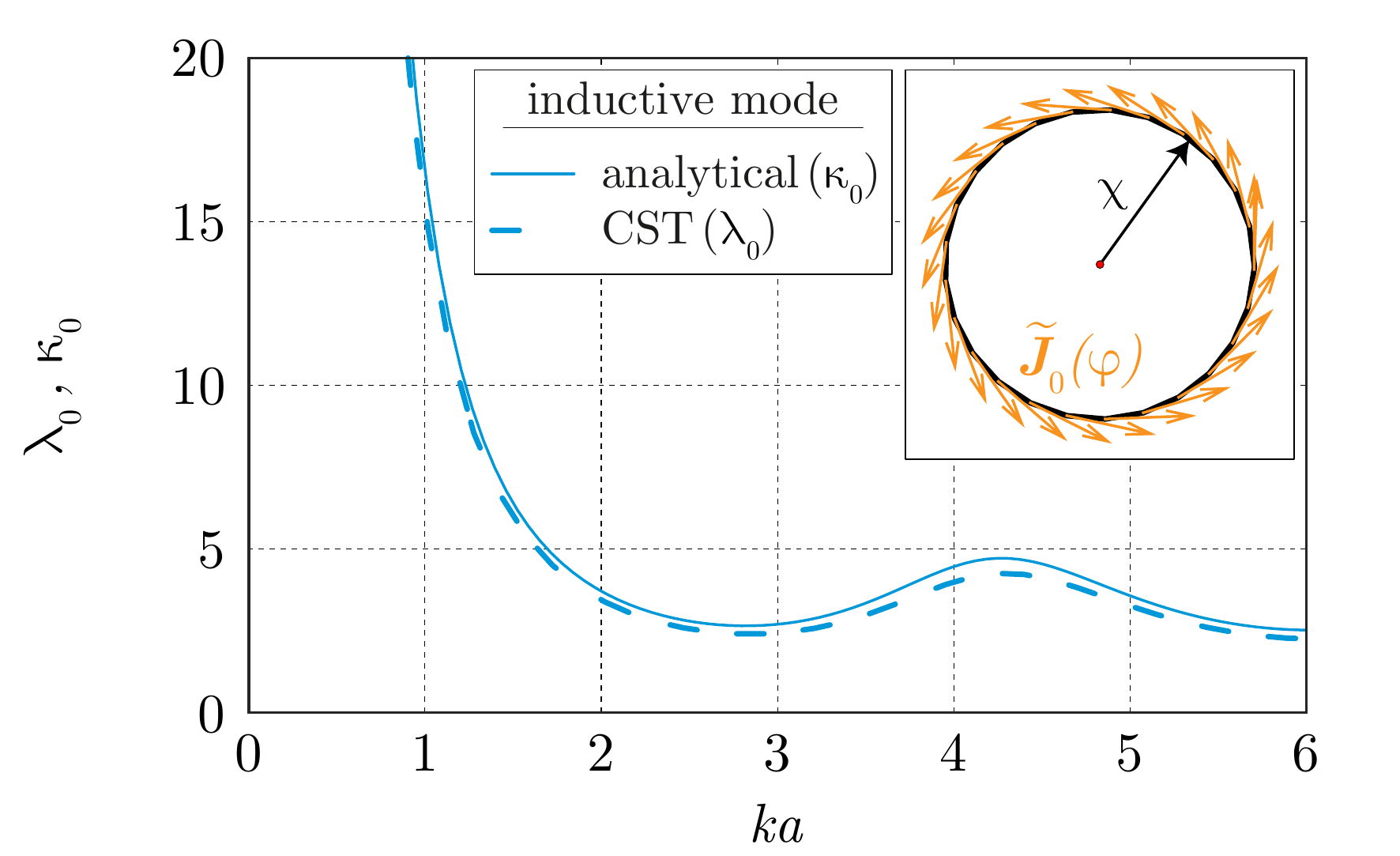}
\caption{Radiation quotient $\kappa_0$ for the uniform mode of loop. A comparison with CM eigenvalue $\lambda_0$ from CST-MWS (1258 triangle elements) is also shown. The uniform mode does not resonate in any given frequency range. However, it can be expected to resonate at extremely high values of $ka$. This behaviour will be closely investigated later.}
\label{fig_loop_fig1}
\end{figure}

The pure inductive character ($\kappa_0>0$) can be clearly seen in Fig.~\ref{fig_loop_fig1}. The agreement between (\ref{TCM_loop1_Ex2}) and $\lambda_0$ obtained by CST is reasonably good as the current is uniquely defined and does not change with frequency. The minor difference is caused by two slightly different models: the reduced kernel with equivalent radius $\myradius/200$ has been utilized to deal with the singularity during the evaluation of (\ref{TCM_loop1_Ex2}), while the thin-strip loop has been calculated in the CST.

Uniform modes do not contribute significantly to far field, but they are important when evaluating near field, input impedance and stored energies.

\section{On the utilization of the analytical functional}
\label{Fun_examples}

The usefulness of the analytical functional (\ref{TCMfcnAJ}) is investigated in a series of examples involving two surface bodies, a cylinder and a spherical shell. It is important to note that the purpose of this section is to demonstrate the potential applications and not to provide a comprehensive treatment of all issues mentioned.

\subsection{Can inductive modes resonate?}
\label{Fun_EX1}

The first example deals with the same topology introduced in Section~\ref{CS_loop} the only difference being the variable height $h$ of the loop. For a significant height, we obtain a cylinder and we need to integrate in $z$-dimension as well. As mentioned already, the uniform mode can occur on the loop-like topology and it is often claimed that this inductive mode, i.e., a mode with \mbox{$\lambda_n \rightarrow \infty$} for \mbox{$ka \rightarrow 0$}, cannot resonate \cite{MartaEva_TheTCMRevisited}. This question can easily be investigated using tools presented in this paper.

The same current (\ref{TCM_loop1_Ex1}) is assumed for the cylindrical shell. Both the uniform (\ref{TCM_loop1_Ex1A}) and the Maxwellian surface current distribution
\begin{equation}
\label{TCM_cylinder2A}
K_0 \left(z\right) = \frac{2}{\pi \sqrt{h^2 - 4 z^2}},
\end{equation}
were tested. In both cases the current was normalized as
\begin{equation}
\label{TCM_cylinder3}
\int\limits_{-\frac{h}{2}}^{\frac{h}{2}} K_0 \left(z\right) \D{z} = 1.
\end{equation}
The results presented in this section were quite insensitive to the choice of (\ref{TCM_loop1_Ex1A}) or (\ref{TCM_cylinder2A}), thus, the distribution physically closer to reality (\ref{TCM_cylinder2A}) was used.

It can easily be seen that the analytical current (\ref{TCM_loop1_Ex1}) has no charge, i.e.,
\begin{equation}
\label{TCM_cylinder4}
\nabla \cdot \widetilde{\boldsymbol{J}}_0 \left(\varphi,r,z\right) = \frac{1}{r} \frac{\partial \widetilde{J_\varphi}}{\partial\varphi} = 0.
\end{equation}
As a consequence, the $\phi\rho_0^\ast$ terms in (\ref{TCMfcnAJ}) are identically zero, which, in conjunction with (\ref{TCM_eq6B}), immediately leads to \mbox{$\We^0 = 0$}. Inspecting (\ref{TCMfcnAJ}), it seems that such a current cannot resonate, however this is only true when \mbox{$\Wm^0$} is always positive, which is not the case here.

\begin{figure}[]
\centering
\includegraphics[width=\figwidth cm]{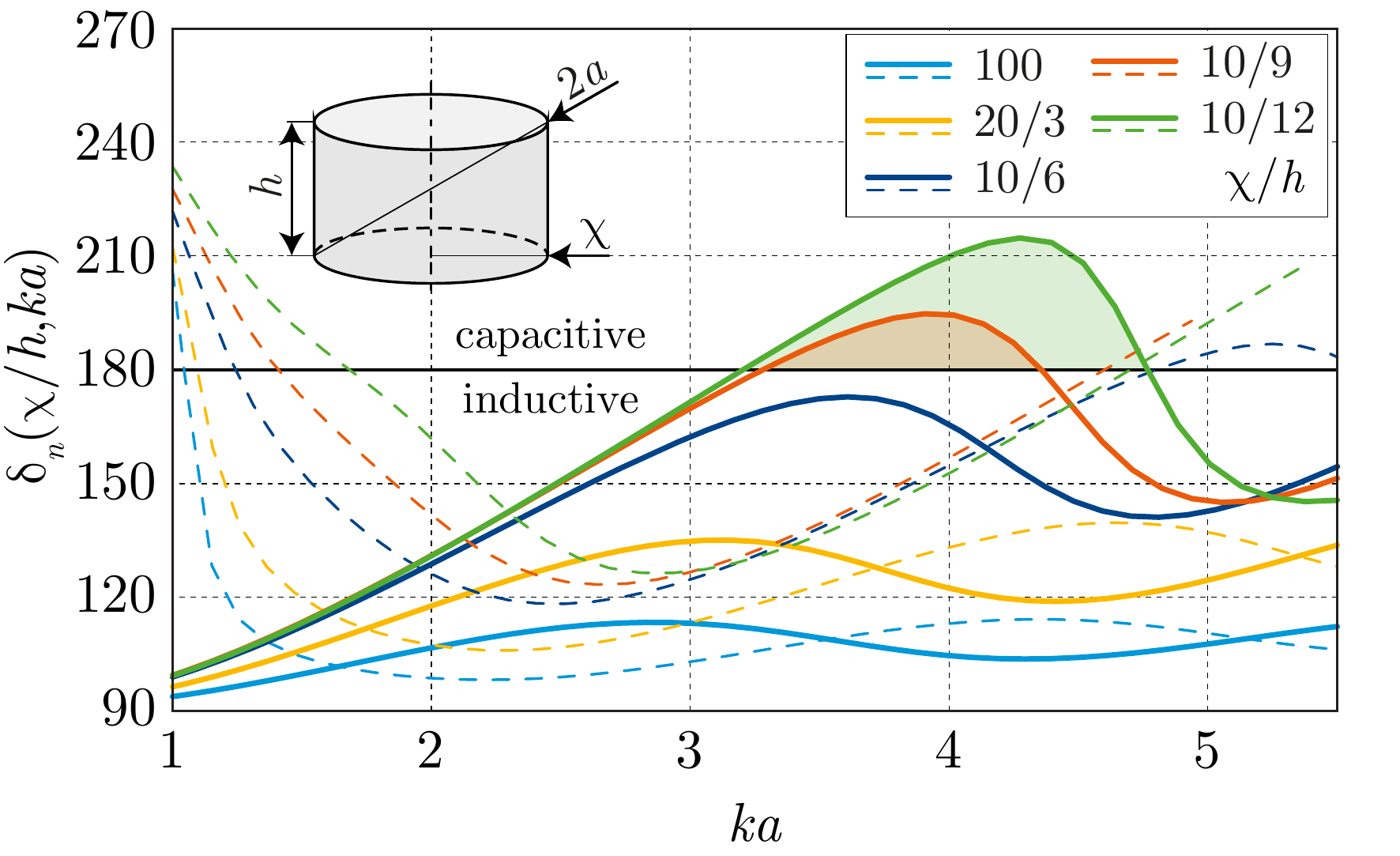}
\caption{Characteristic numbers of a PEC cylinder depicted in terms of characteristic angles $\charnum_n$ as a function of $ka$ and radius to height ratio. The exact dimensions of the cylinder are shown in the inset. The capacitive modes are depicted by dashed lines, whereas the inductive modes are depicted by solid lines. It can be seen that the modes for \mbox{$\myradius/h = \left\{10/9,10/12\right\}$} cross the resonance line \mbox{$\delta = 180\degree$} at \mbox{$ka \approx 3.2$.}}
\label{fig_cylinder1}
\end{figure}

\begin{figure}[]
\centering
\includegraphics[width=\figwidth cm]{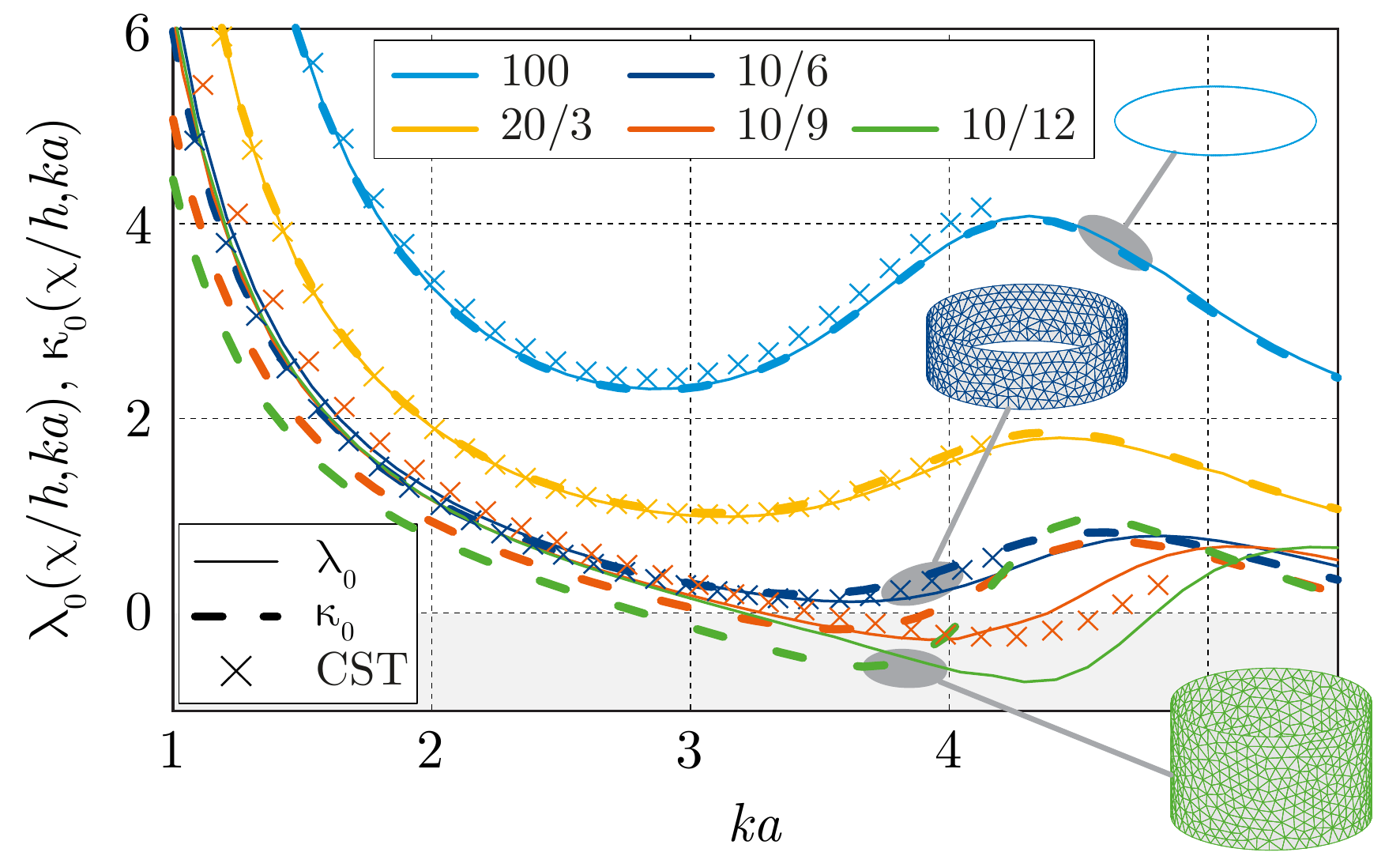}
\caption{Characteristic numbers of uniform mode of the PEC cylinder from Fig.~\ref{fig_cylinder1} are depicted as a function of $ka$. The dimensions are the same as in Fig.~\ref{fig_cylinder1}. For comparison purposes, the eigenvalues $\lambda_0$ calculated in AToM are compared with CST-MWS (cross markers) and with analytical evaluation of $\kappa_0$ given by (\ref{TCM_cylinder5}) and (\ref{TCM_loop1_Ex1A}) (dashed lines). The overall agreement is good, however, the evaluation of the analytical current varies from true characteristic mode for higher $ka$ and higher $\myradius/h$. Even in such cases the qualitative behaviour is the same -- the uniform mode can resonate.}
\label{fig_cylinder2}
\end{figure}

The uniform mode for a cylinder of various $\myradius / h$ is depicted in Fig.~\ref{fig_cylinder1} in terms of eigenangles \cite{Newman_SmallAntennaLocationSynthesisUsingCharacteristicModes}
\begin{equation}
\label{TCM_cylinder6}
\delta_n = \frac{180}{\pi} \left(\pi - \mathrm{atan} \left(\lambda_n\right) \right).
\end{equation}
The characteristic eigenangles normalize the eigenvalues and indicate the electromagnetic behaviour of CMs. Modes are capacitive for \mbox{$\delta_n > 180\degree$}, inductive for \mbox{$\delta_n < 180\degree$} and resonate for \mbox{$\delta_n = 180\degree$}. We can see in Fig.~\ref{fig_cylinder1} that the uniform mode of the sufficiently tall cylinder can cross the resonance even if it lacks $\We$ energy (charge). This observation is verified in Fig.~\ref{fig_cylinder2} in which the eigenvalues were calculated using the AToM package \cite{atom} (solid lines), in CST-MWS (cross markers) and finally evaluated according to (\ref{TCM_loop1_Ex1A}) with (\ref{TCM_loop1_Ex1}) substituted
\begin{equation}
\label{TCM_cylinder5}
\kappa_0 = \frac{\displaystyle\int\limits_0^{2\pi} \int\limits_{-\frac{h}{2}}^{\frac{h}{2}} \int\limits_{-\frac{h}{2}}^{\frac{h}{2}} K_0 \left(z_1\right) K_0 \left(z_2\right) \cos\left(\varphi\right) \frac{\cos\left(k R\right)}{R} \D{z_1} \D{z_2} \D{\varphi}}{\displaystyle\int\limits_0^{2\pi} \int\limits_{-\frac{h}{2}}^{\frac{h}{2}} \int\limits_{-\frac{h}{2}}^{\frac{h}{2}} K_0 \left(z_1\right) K_0 \left(z_2\right) \cos\left(\varphi\right) \frac{\sin\left(k R\right)}{R} \D{z_1} \D{z_2} \D{\varphi}}.
\end{equation}
where \mbox{$R = \sqrt{2 \myradius^2 \left(1 - \cos\left(\varphi\right)\right) + \left(z_1 - z_2\right)^2}$} and the axial symmetry of the cylinder have been utilized as in (\ref{TCM_loop1_Ex2}) in order to reduce one of integrals in $\varphi$ direction.

\begin{figure}[]
\centering
\includegraphics[width=\figwidth cm]{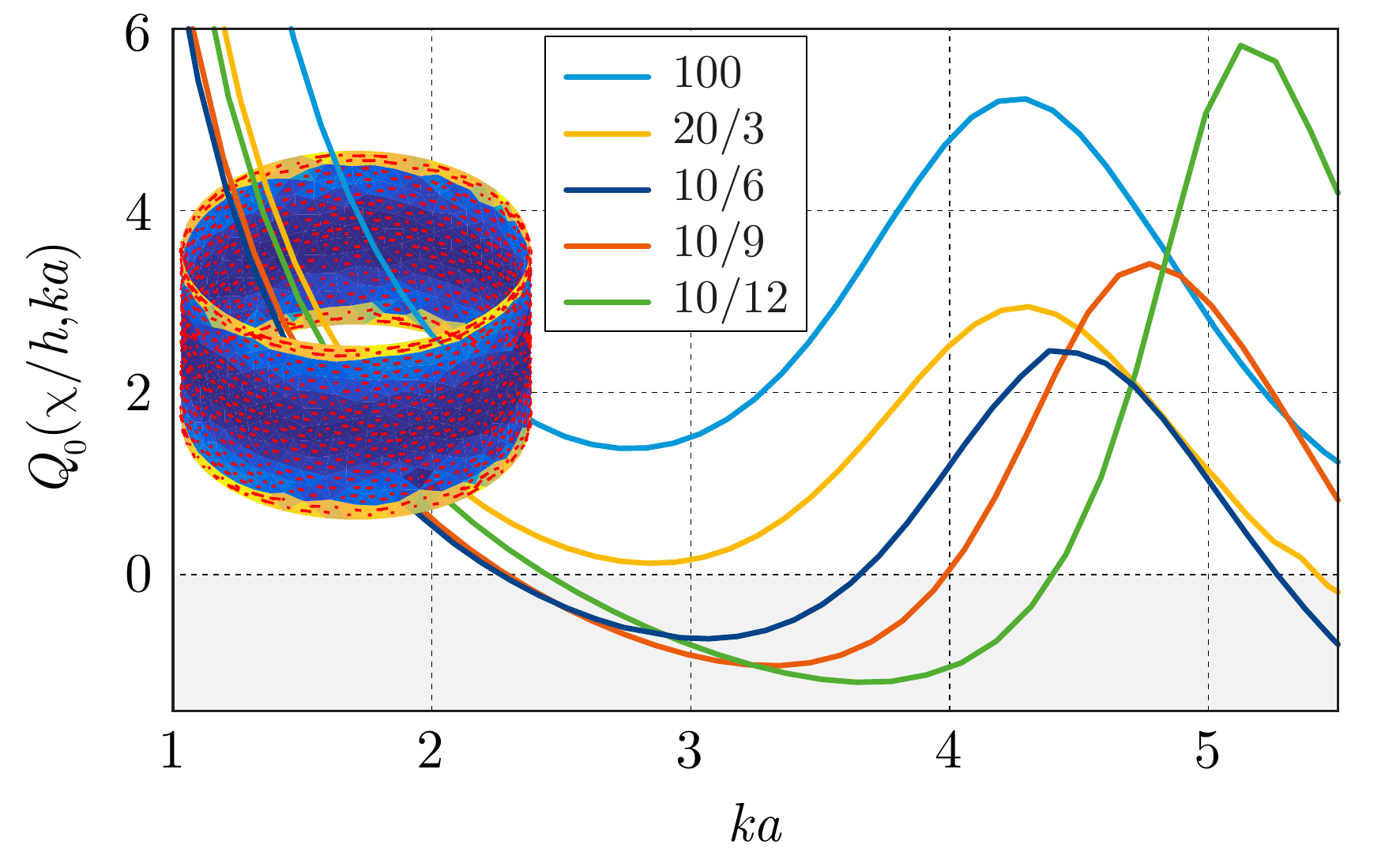}
\caption{Quality factor~$Q$ of uniform modes from Fig.~\ref{fig_cylinder2} for the same dimensions of the PEC cylinder. Thanks to the additional term $\Wrad$ \cite{Vandenbosch_ReactiveEnergiesImpedanceAndQFactorOfRadiatingStructures}, quality factor~$Q$ is negative not only for modes with \mbox{$\myradius/h = \left\{10/9,10/12\right\}$}, but also for \mbox{$\myradius/h = 10/6$} (blue curve). The distribution of the current density on the cylinder is depicted in the inset.}
\label{fig_cylinder3}
\end{figure}

The fact that the uniform mode can resonate, even when \mbox{$\We^0 = 0$}, clearly indicates that the term $\Wm^0$ can be negative. Finally, using the formula for modified stored energy (\ref{TCM_storedEnergyN1}) from \cite{Vandenbosch_ReactiveEnergiesImpedanceAndQFactorOfRadiatingStructures} 
\begin{equation}
\label{TCM_cylinder7}
\WSTO = \Wm + \We + \Wrad
\end{equation}
and evaluating it according to formulas (63) and (64) in \cite{Vandenbosch_ReactiveEnergiesImpedanceAndQFactorOfRadiatingStructures}, we obtain the values of quality factor~$Q$. The results are depicted in Fig.~\ref{fig_cylinder3}. The uniform mode on the tall cylinder has a negative value of modified stored energy $\WSTO$, which means that \mbox{$\Wm^0 < \Wrad$} since \mbox{$\We^0 = 0$}. This is equivalent to the negative slope of eigenvalue $\lambda_0$ in (\ref{TCM_storedEnergyN10}) and both observations lead to the negative value of the quality factor~$Q$.

The same behaviour has already been described in \cite{GustafssonCismasuJonsson_PhysicalBoundsAndOptimalCurrentsOnAntennas_TAP}, and, so far, only loop-like, divergence-free currents were found, which supports the reasoning in \cite{GustafssonCismasuJonsson_PhysicalBoundsAndOptimalCurrentsOnAntennas_TAP}. Using another kind of analysis, the characteristic modes, we hypothesize that the problem is caused by extraordinary uniform currents with \mbox{$\We \approx 0$} which, in reality, cannot exist independently (it can be shown that there is no realistic feeding that can excite only $\boldsymbol{J}_0$ mode).

\subsection{Numerical analysis of CMs as GEP -- Benchmarks utilizing a spherical shell}
\label{Fun_EX3}

Bearing in mind the results of the previous sections, we can perform a series of benchmarks, employing the knowledge of characteristic modes and numbers in analytical form. To do this, we need to find a scatterer whose characteristic modes are known analytically. The perfect candidate is a spherical shell, whose characteristic fields coincide with properly normalized spherical harmonics \cite{Garbacz_TCMdissertation}. The characteristic numbers can be evaluated analytically if the characteristic currents are substituted into (\ref{TCMfcnAJ}). This becomes of interest when dealing with the numerical solvers which are encumbered with rounding (and other) numerical errors.

\begin{figure}[]
\centering
\includegraphics[width=\figwidth cm]{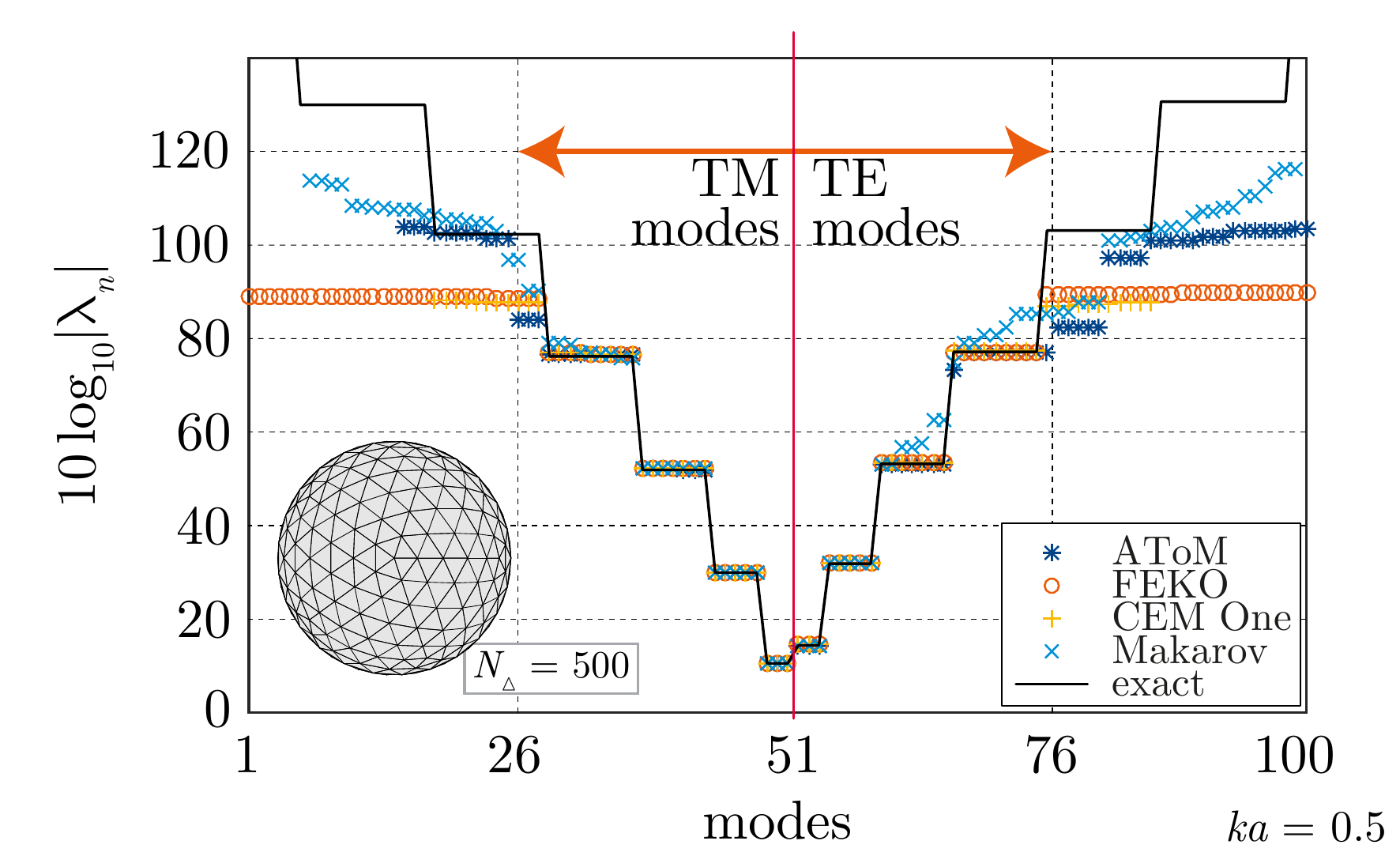}
\caption{The characteristic numbers $\lambda_n$ of the spherical shell at \mbox{$ka = 0.5$} are depicted for four numerical solvers. The sphere was discretized into $N_\triangle = 500$ triangles. Modes up to TM$_{4l}$ and TE$_{4l}$ have been found correctly, including their degeneracy. The magnitude of all modes has been limited to values $\left|\lambda_\mathrm{max}\right| = 10^{11}$ with the characteristic modes being sorted according to their magnitude. Characteristic numbers on the left side originally had negative values, whereas numbers on the right side had positive values. The exact values of the characteristic numbers are depicted by the solid black line.}
\label{fig_sphere1}
\end{figure}

\begin{figure}[]
\centering
\includegraphics[width=\figwidth cm]{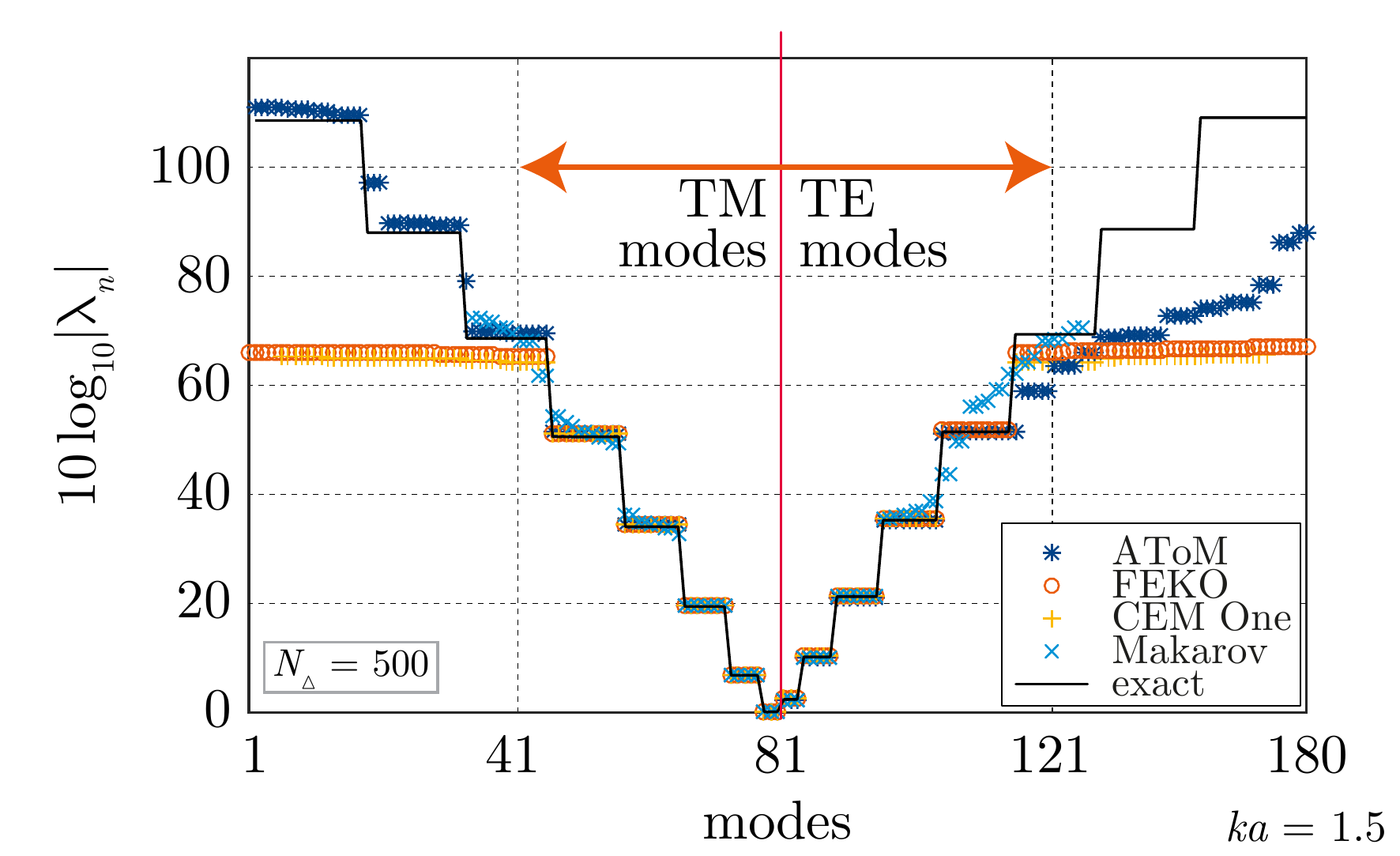}
\caption{The characteristic numbers $\lambda_n$ for the spherical shell at \mbox{$ka = 1.5$}. The curves represent the same quantities as in Fig.~\ref{fig_sphere1}, including the results processing. Compared to Fig.~\ref{fig_sphere1} more than two times the number of modes have been found correctly since all modes are closer to their resonance.}
\label{fig_sphere2}
\end{figure}

\begin{figure}[]
\centering
\includegraphics[width=\figwidth cm]{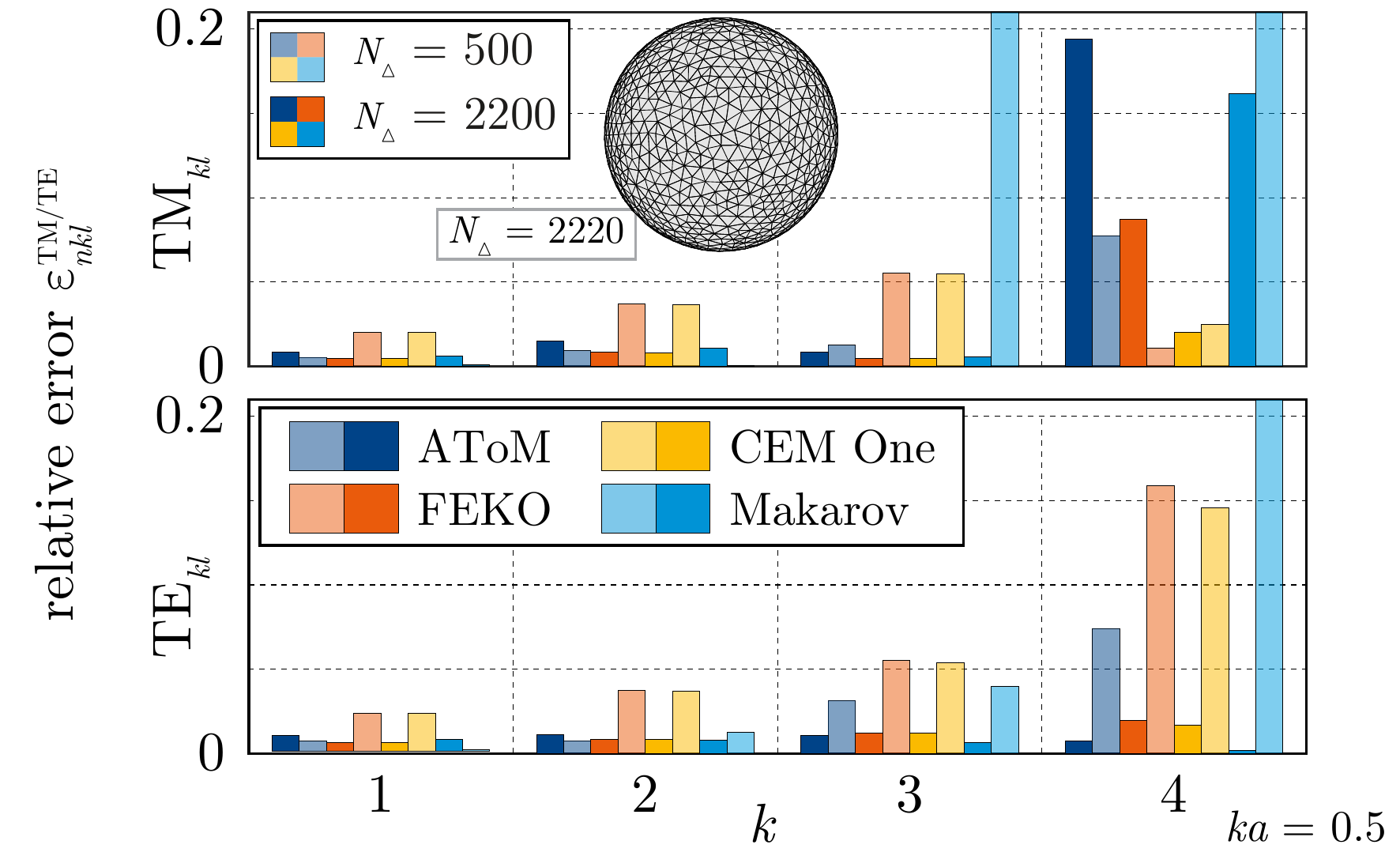}
\caption{Comparison of the relative errors of the first four TM and TE characteristic modes of a spherical shell at \mbox{$ka = 1/2$}. The relative error is evaluated with respect to the analytically evaluated characteristic numbers, while arithmetic mean of the characteristic numbers of all degenerated modes has been calculated and depicted by error bars. The selected software packages are differentiated by various colours, with mesh densities depicted by dark and light tints.}
\label{fig_sphere3}
\end{figure}

The characteristic numbers $\lambda_n$, obtained using decomposition (\ref{TCM_eq4B}) of the impedance matrices $\mathbf{Z}$ from different commercial and in-house packages, are compared with exact radiation coefficients $\kappa_{kl}^\mathrm{TE/TM}$ calculated via (\ref{TCMfcnAJ}) for spherical harmonics $\boldsymbol{J}_{kl}^\mathrm{TE/TM}$ \cite{Lebedev_SpecialFunctions}, see Fig.~\ref{fig_sphere1} and Fig.~\ref{fig_sphere2}, respectively. The software packages have been used to generate impedance matrices and, in all cases, the eigen-decomposition has been performed in Matlab. The exact characteristic numbers are depicted by solid black lines and constitute known references. It can be seen that even state-of-the-art commercial simulators are capable of finding only the first four TM$_{kl}$ and TE$_{kl}$ modes. This is caused mainly by the $2k+1$ degeneracy (the number of degenerated modes increases rapidly) and by the limited (double) numerical precision. Surprisingly, the number of well-defined modes is not influenced by the number of discretization triangles $N_\Delta$. On the contrary, the relative error between analytically and numerically calculated characteristic numbers is a function of $N_\Delta$ which is confirmed by Fig.~\ref{fig_sphere3}. While the relative error of dominant TM$_{1l}$ and TE$_{1l}$ modes is a few percentage points, it quickly reaches about 10\,\% for groups of TM$_{4l}$ and TE$_{4l}$ modes. The overall results, presented in Figs.~\ref{fig_sphere1}--\ref{fig_sphere3}, favour the in-house Matlab tool AToM. However, FEKO \cite{feko} and CEM~One \cite{ESI_VisualCEM} packages reach comparable results. The routines available for free in \cite{Makarov_AntennaAndEMModelingWithMatlab} suffer from non-symmetry of produced impedance matrices. This issue can be resolved manually during pre-processing to reduce the relative error significantly. Notice that CST is not depicted since the impedance matrices cannot be acquired.

Other tests, e.g., those involving modal currents, can be performed as well. For example, the numerically calculated characteristic modes on the spherical shell can be compared with their analytical forms via
\begin{equation}
\label{TCM_Compar1}
\epsilon_{nkl} = \left\langle \boldsymbol{J}_n , \boldsymbol{J}_{kl}^\mathrm{TE/TM} \right\rangle.
\end{equation}
However, that study goes beyond the scope of this paper.

\section{Conclusion}
The paper discusses specific advances of the theory of characteristic modes as introduced by Garbacz, Harrington and Mautz, but expressed here in terms of a particular functional, which is minimized by eigencurrents. This novel formula provides a different perspective on characteristic mode decomposition.

The usefulness of the functional is illustrated by three canonical examples: a dipole, two closely spaced dipoles and a loop. It was shown that the functional formulation is better suited to be analysed than the original formulation because there is no impedance matrix involved. A deeper investigation of the modes on a dipole reveals the limitations of the approximation of the zero-order current distribution expressed as a $\sin$ function.

Knowledge of the analytical functional is important for a few significant topics dealing with various issues of antenna analysis and design. In particular, any method of moment code or characteristic mode solver can be benchmarked using analytical results for a spherical shell.

\appendix[Relationship between $Q_n$ and $Q_{\mathbf{X},n}$] 
\label{app1}

The purpose of this appendix is to derive (\ref{TCM_storedEnergyN12}). To simplify the underlying mathematical nomenclature, the derivation is done for all quantities in their matrix forms. First, modal quality factor Q (\ref{TCM_storedEnergyN10}) is expressed in terms of characteristic currents using the matrix form of (\ref{TCM_eq5}) as
\begin{equation}
\label{TCM_App1}
Q_n = \frac{\omega}{2} \frac{\partial}{\partial\omega}\left(\frac{\Ivec_n^\herm \mathbf{X}\Ivec_n}{\Ivec_n^\herm\mathbf{R}\Ivec_n}\right),
\end{equation}
then the differentiation is performed
\begin{equation}
\label{TCM_App2}
Q_n = Q_{\mathbf{X},n} + \frac{\displaystyle\omega \frac{\partial\Ivec_n^\herm}{\partial\omega} \mathbf{X} \Ivec_n}{\Ivec_n^\herm \mathbf{R} \Ivec_n} - \frac{1}{2} \frac{\displaystyle\Ivec_n^\herm \mathbf{X} \Ivec_n \omega \frac{\partial}{\partial\omega} \left(\Ivec_n^\herm \mathbf{R} \Ivec_n\right)}{\left( \Ivec_n^\herm \mathbf{R} \Ivec_n \right)^2},
\end{equation}
in which the quality factor (\ref{TCM_storedEnergyN12}) has been substituted and the following identity has been employed
\begin{equation}
\label{TCM_App3}
\omega \frac{\partial\Ivec_n^\herm}{\partial\omega} \mathbf{X} \Ivec_n + \Ivec_n^\herm \mathbf{X} \omega\frac{\partial\Ivec_n}{\partial\omega} = 2 \omega \frac{\partial\Ivec_n^\herm}{\partial\omega} \mathbf{X} \Ivec_n
\end{equation}
since the modal currents are supposed to be purely real. Then, the RHS of (\ref{TCM_eq4B}) is substituted into the last two terms on the RHS of (\ref{TCM_App2}), which yields
\begin{equation}
\label{TCM_App4}
Q_n = Q_{\mathbf{X},n} + \lambda_n \left( \frac{\displaystyle\omega \frac{\partial\Ivec_n^\herm}{\partial\omega} \mathbf{R} \Ivec_n}{\Ivec_n^\herm \mathbf{R} \Ivec_n} - \frac{1}{2} \frac{\displaystyle \omega \frac{\partial}{\partial\omega} \left(\Ivec_n^\herm \mathbf{R} \Ivec_n\right)}{\Ivec_n^\herm \mathbf{R} \Ivec_n} \right).
\end{equation}
Finally, performing the differentiation in the last term on the RHS of (\ref{TCM_App4}) and using identity (\ref{TCM_App3}), we get
\begin{equation}
\label{TCM_App5}
Q_n = Q_{\mathbf{X},n} - \lambda_n \frac{\displaystyle \Ivec_n^\herm \omega \frac{\partial \mathbf{R}}{\partial\omega} \Ivec_n}{2 \Ivec_n^\herm \mathbf{R} \Ivec_n} = Q_{\mathbf{X},n} - \lambda_n Q_{\mathbf{R},n}.
\end{equation}

\section*{Acknowledgement}
The authors would like to thank Lukas~Jelinek and Leslie~Ryan for their valuable comments and Mats~Gustafsson for a fruitful discussion which led to the discovery of crucial ideas presented in the paper. The authors are also grateful for the opportunity to use the method of moment code written by Vladimir Sedenka. They would also like to thank the three anonymous reviewers whose remarks improved the clarity of this paper.

\ifCLASSOPTIONcaptionsoff
  \newpage
\fi


\bibliographystyle{IEEEtran}
\bibliography{references_LIST}
\begin{biography}[{\includegraphics[width=1in,height=1.25in,clip,keepaspectratio]{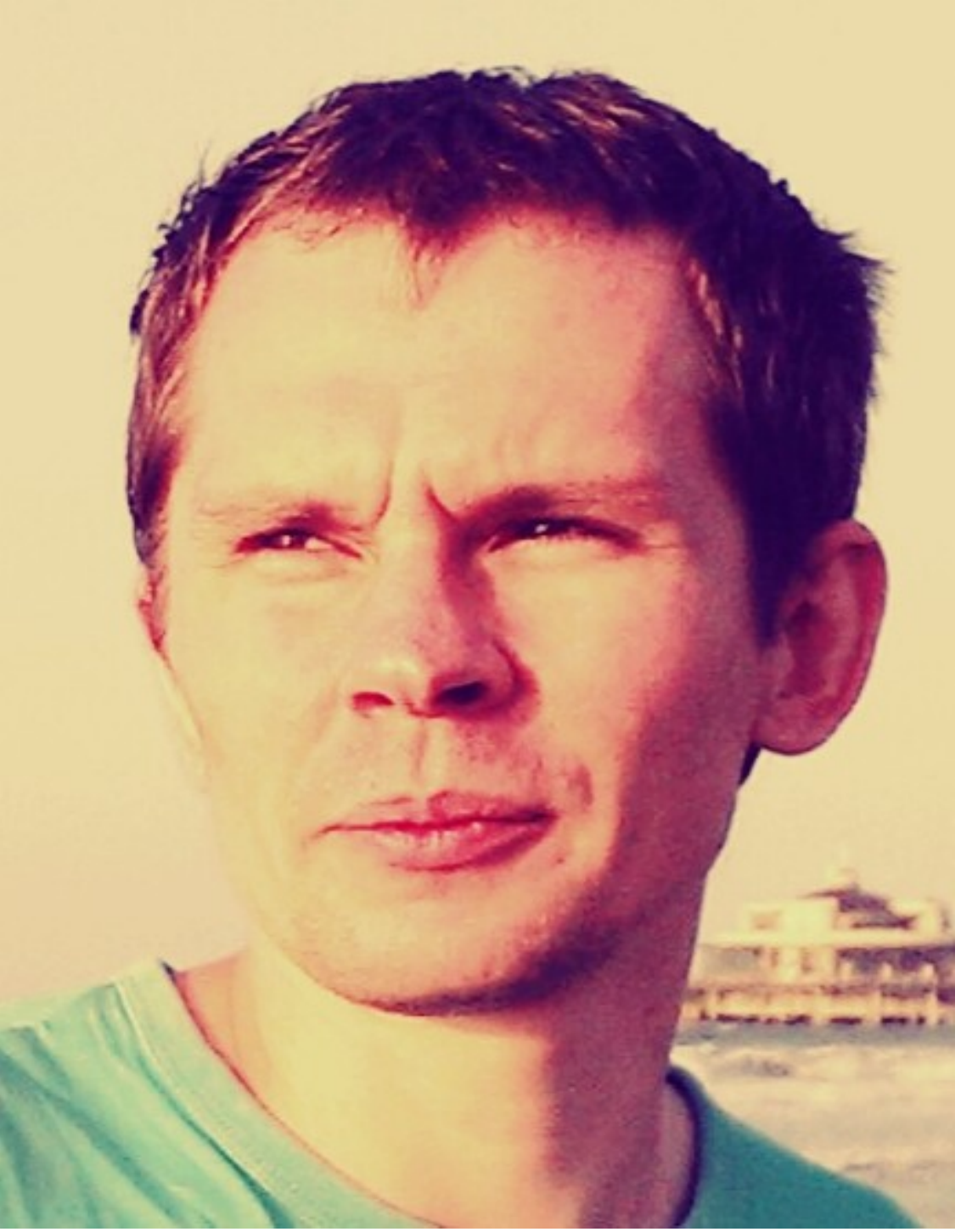}}]{Miloslav Capek}
(S'09, M'14) received his M.Sc. degree in Electrical Engineering from the Czech Technical University, Czech Republic, in 2009, and his Ph.D. degree from the same University, in 2014. Currently, he is a researcher with the Department of Electromagnetic Field, CTU-FEE.
	
He leads the development of the AToM (Antenna Toolbox for Matlab) package. His research interests are in the area of electromagnetic theory, electrically small antennas, numerical techniques, fractal geometry and optimization. He authored or co-authored over 45 journal and conference papers.

Dr. Capek is member of Radioengineering Society, regional delegate of EurAAP, and Associate Editor of Radioengineering.

\end{biography}
\begin{biography}[{\includegraphics[width=1in,height=1.25in,clip,keepaspectratio]{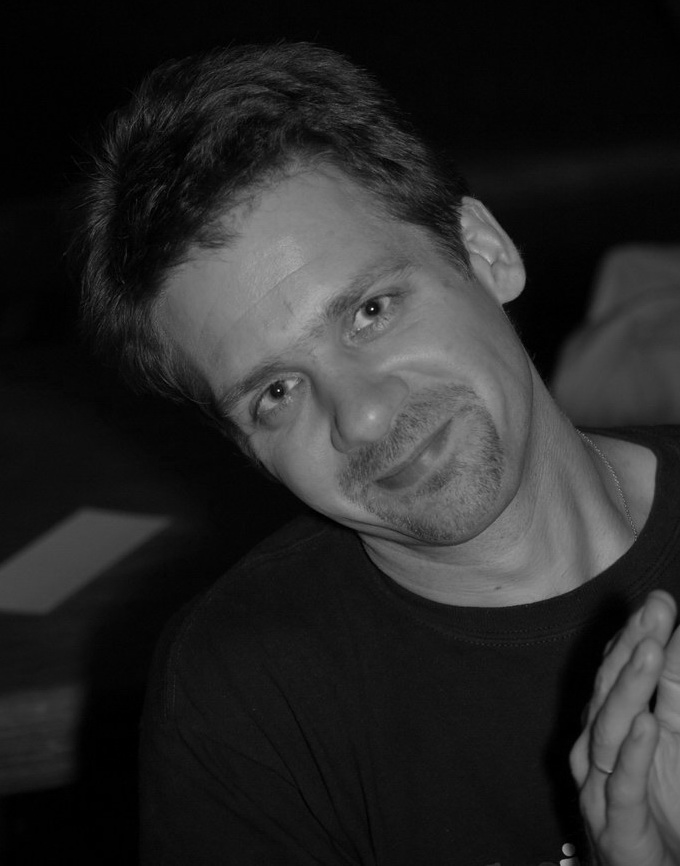}}]{Pavel Hazdra}
(M'03) received the M.Sc. and Ph.D. degree in electrical engineering from the Czech Technical University in Prague, Czech Republic, in 2003 and 2009, respectively. Since 2012 he is an associate professor with the Department of Electromagnetic Field at the CTU in Prague. He authored or co-authored more than 15 journal and 20 conference papers. His research interests are in the area of EM/antenna theory, electrically small antennas, reflector antennas and their feeds and antennas for radioamateur purposes.

Dr. Hazdra is member of the board of Radioengineering Society.

\end{biography}

\begin{biography}[{\includegraphics[width=1in,height=1.25in,clip,keepaspectratio]{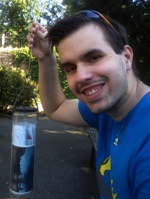}}]{Michal Masek}
received the M.Sc. degree in electrical engineering from the Czech Technical University in Prague, Czech Republic, in 2015. He is now working towards his Ph.D. degree in the area of behavior of small antennas close to large objects.

\end{biography}

\begin{biography}[{\includegraphics[width=1in,height=1.25in,clip,keepaspectratio]{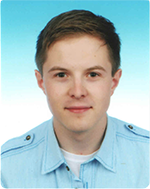}}]{Vit Losenicky}
received the M.Sc. degree in electrical engineering from the Czech Technical University in Prague, Czech Republic, in 2016. He is now working towards his Ph.D. degree in the area of electrically small antennas.

\end{biography}
\end{document}